\documentstyle[12pt,aasms4]{article}
\newcommand{\kms}{\mbox{${\rm km~s}^{-1}$}}
\renewcommand{\bv}{\mbox{$(B-V)$}}
\newcommand{\vsini}{\mbox{$v\,\!\sin\,\!i$}}
\newcommand{\lR}{\mbox{$\log R_{\rm HK}^\prime$}}
\doublespace
\begin{document}
\title{THE PROBLEM OF HIPPARCOS DISTANCES TO OPEN CLUSTERS. II.
CONSTRAINTS FROM NEARBY FIELD STARS\footnotemark[1]}
\footnotetext[1]{Based on data from the ESA Hipparcos astrometry satellite.}

\author{David R. Soderblom, Jeremy R. King}
\affil{Space Telescope Science Institute\\
3700 San Martin Drive, Baltimore MD 21218\\
email: soderblom@stsci.edu, jking@stsci.edu}

\author{Robert B. Hanson, Burton F. Jones}
\affil{University of California Observatories/Lick Observatory\\
Board of Studies in Astronomy and Astrophysics\\
University of California, Santa Cruz CA 95064\\
email: hanson@ucolick.org, jones@ucolick.org}

\author{Debra Fischer}
\affil{Dept. of Physics and Astronomy\\
San Francisco State University, San Francisco CA 94132\\
email: fischer@stars.sfsu.edu}

\author{John R. Stauffer}
\affil{Harvard-Smithsonian Center for Astrophysics\\
60 Garden Street, Cambridge MA 02138\\
email: stauffer@cfa.harvard.edu}

\and

\author{Marc H. Pinsonneault}
\affil{Astronomy Department, Ohio State University,\\
174 West 18th Avenue, Columbus OH 43210\\
email: pinsono@astronomy.ohio-state.edu}

\begin{abstract}
This paper examines the discrepancy between distances to nearby open clusters
as determined by parallaxes from Hipparcos compared to traditional main
sequence fitting.  The biggest difference is seen for the Pleiades, and
our hypothesis is that if the Hipparcos distance
to the Pleiades is correct, then similar subluminous ZAMS stars should exist
elsewhere, including the immediate solar neighborhood.  We examine a
color-magnitude diagram of very young and nearby solar-type stars and show
that none of them lie below the traditional ZAMS, despite the fact that the
Hipparcos Pleiades parallax would place its members 0.3 magnitude below
that ZAMS\@.  We also present analyses and observations of solar-type stars
that do lie below the ZAMS and show that they are subluminous because of low
metallicity and that they have the kinematics of old stars.

\end{abstract}

\section{Distances to Open Clusters}

The results of the Hipparcos mission have recently appeared
(\cite{ESA97}, 1997),
and they provide unprecedented astrometric precision and accuracy for a very
large sample of stars.  Analyses of these results are just beginning, of
course, but to us some of the most intriguing Hipparcos measurements
are those of nearby open clusters, such as the Hyades, Pleiades, Praesepe,
and $\alpha$ Persei.

Open clusters are critical laboratories for testing stellar evolution models
since they provide large samples of stars of a single age and composition
(as near as we can tell, anyway) over a broad range of mass.  Those models are
calibrated against the Sun, the one star for which we know fundamental
properties with very high accuracy.  Thus we construct stellar models, adjust
them to match the Sun, and then test them against open clusters because those
clusters have near-solar composition, making it possible to work
differentially.  Having passed those tests, we have some confidence the models
can then be applied to significantly different conditions, such as globular
clusters, which are vital for establishing the cosmic distance scale.

This paper and the companion paper by \cite{Pins98} (1998) are
motivated by concern over the accuracy of the Hipparcos results.
Nearly all the Hipparcos cluster distances disagree with
conventionally-determined values, although in most cases the differences do
not conflict with the estimated uncertainties.  But the Pleiades is an
especially egregious case.  The Hipparcos estimates of the Pleiades parallax
range from 8.54 to 8.65 mas, depending on the solution used: Robichon gets
$8.54\pm0.22$ (see Table XXVI of \cite{FVL97} 1997); \cite{Mer97} (1997) get
$8.60\pm0.24$; \cite{vL97} (1997) quote $8.61\pm0.23$ as their solution A
[this value also appears in \cite{vLE97} (1997) and \cite{FVL97} (1997)]; and
\cite{vL97} (1997) cite $8.65\pm0.24$ as their solution B\@.
These correspond to a distance of about 116 pc or a distance modulus of 5.33
magnitudes.  Traditional determinations of the
cluster's distance (e.g., \cite{VdB84} 1984; \cite{SSHJ} 1993,
hereafter SSHJ) are based on comparing the cluster's main sequence to that of
nearby stars, and these lead to a distance modulus of about 5.6.  The same
value of 5.6 has been derived by fitting a spectroscopic binary to isochrones
(\cite{Gia95} 1995).  \cite{BF90} (1990) show that the Pleiades has [Fe/H]
= $-0.034\pm0.024$, i.e., it has essentially solar metallicity.  Thus the
Hipparcos results suggest that Pleiades members are about 0.3 magnitude
fainter than we have thought up to now.  Can these different estimates be
reconciled?  Can a Zero-Age Main Sequence (ZAMS) star with solar
metallicity be 30\% fainter than our current models predict?  These are the
essential questions that we address here.

The Hipparcos parallax of van Leeuwen \& Hansen Ruiz is based on measurements
of 54 Pleiades members,
ranging in $m_V$ from about 3 to 11, so it represents a good cross-section
of the cluster.  Hipparcos observations are reduced to an absolute
reference frame, but the measurements are correlated within a limited region
of the sky as the satellite swept out great circles.  These correlated
measures have been corrected for (\cite{vL97} 1997) as part
of the effort to reduce all the Hipparcos observations in a consistent
and systematic way.  Reconciling the Hipparcos distance with the traditional
estimate would imply systematic errors larger than the quoted uncertainties.
There is, therefore, no obvious reason to believe the
Hipparcos distance to the Pleiades contains a systematic error that is
large enough to bring it into accord with the traditional distance.

The traditional distance measure, on the face of it, appears to be just as
sound.  It is based on comparing a Pleiades color-magnitude diagram (CMD)
-- corrected for a small amount of reddening -- to a CMD created from nearby
stars with large parallaxes, or to a CMD of the Hyades, suitably corrected
for the difference in metallicity.  Theoretical isochrones can also be
converted to observational coordinates using a color calibration, and the
offset between the isochrone and the cluster can be used to infer the
distance modulus.  This technique is used in the companion paper by
Pinsonneault et al. and yields similar
results.  In this paper we reexamine the comparison of the Pleiades to nearby
stars.  Our hypothesis is that the stars of the Pleiades cannot be completely
unique in our Galaxy and that there must be nearby examples of stars that
share the same unknown stellar physics or unusual parameters that result in
the Pleiades stars being so faint.  It should therefore be possible to find
examples of anomalously-faint ZAMS stars that are so close to the Sun that
errors in parallax cannot account for their faintness.  If no such stars
exist, as we will show, then either we have failed to account for some
fundamental aspects of stellar physics adequately, or there are unappreciated
errors in the Hipparcos parallaxes.

\section{An Observational ZAMS Using Nearby Solar-Type Stars}

We start by showing that nearby solar-type stars that are known to be young
do not lie below the usual ZAMS.  The idea of comparing a cluster main
sequence to one constructed from nearby stars with large parallaxes is not new,
but the nearby stars are of many ages and evolutionary states, which spreads
the apparent main sequence considerably.  The appropriate comparison, of
course, is to very young nearby stars, since the clusters in questions are
essentially ZAMS themselves.

In this case by young we mean very active, as determined from observations of
the \ion{Ca}{2} H and K lines.  Table 1 lists our sample.  The northern stars
have been observed as part of the Mount Wilson survey of chromospheric
emission in late-type dwarfs (\cite{VP80} 1980; \cite{Sod85} 1985; \cite{SM93}
1993), from which we have taken the $R_{\rm HK}^\prime$ index of HK emission.
To the extent they have been measured, these stars have metallicities near
solar (\cite{Cay92} 1992).  The photometry of the northern stars is from
\cite{MM94} (1994).  We divided these northern stars into two subsets.  The
first consists of the most active of the stars, those with \lR $>-4.40$, to
which we added a few others which are slightly less active but which are so
well studied that there is no ambiguity about their youth (HD 39587 =
$\chi^1$ Ori is an example).  The second subset of northern stars is also
active, but not as much so or not as well-studied; they have \lR\ values from
$-4.41$ to $-4.44$.  We have also included some southern stars from the HK
survey of \cite{Hen96} (1996) that have \lR\ values from $-4.20$ to $-4.40$;
that paper provides the photometry.

The parallaxes in Table 1 are from Hipparcos (\cite{ESA97}, 1997).
We kept only those stars with $\sigma_\pi/\pi \la 0.1$ so that parallax error
could not accidentally place a star significantly below the ZAMS.
We also excluded stars with known companions unless we were confident that
the companion is not influencing the HK observations or the photometry.

Our young stars are shown in Figure 1.
The large dots represent the first subset; i.e., the stars most likely to be
{\it bona fide} ZAMS objects.  The small dots represent the other northern
stars and the open circles are the southern stars.  The solid line is a
theoretical ZAMS from VandenBerg (1997, private communication).  It has been
calibrated to reproduce the solar temperature and luminosity (represented by
the diamond) at the Sun's age, and to fit the M67 cluster main sequence at its
age.  The dashed line is the same ZAMS transformed to the CMD using the
color-temperature relation of \cite{Be79} (1979).  For reference, the
long-dashed line shows the same ZAMS (for 100 Myr age and [Fe/H] = 0.0) used
in the companion paper by Pinsonneault et al.  About half the difference
between the VandenBerg and Pinsonneault isochrones arises in the
color-temperature relations used.  Their zero points are close (the
VandenBerg isochrone is, on average, 0.04 magnitude fainter in the range of
0.5 to 0.9 in \bv), and there is a slight difference in the slopes of the
main sequences.  Differences in the color-temperature relations are a larger
source of uncertainty for the cooler stars, as the increasing difference
between the VandenBerg and Bessell lines indicates.

The theoretical isochrones are clearly an excellent representation of the
observations.  We anticipate finding stars above the ZAMS by modest amounts
because they are photometric binaries, but we note that none of the young
stars falls below the ZAMS\@.  Thus there is no hint in this small sample of
there being any nearby young stars that are 0.3 magnitude below the usual ZAMS.

Figure 2 shows a similar CMD for the Pleiades, taken from SSHJ
and corrected for reddening of 0.04 magnitude in \bv\ and 0.12
magnitude in $V$.  The lines are the same ones as in Figure 1, but displaced
by 5.6 magnitudes.  This comparison shows that different isochrones can
differ from one another and from the cluster by 0.1 magnitude or more for
\bv $\ga0.7$.  The Bessell relation is clearly too blue, while both the
VandenBerg and Pinsonneault isochrones are too red for \bv $\ga0.8$.
Note, however, that these theoretical ZAMS lines deviate from the
Pleiades in the same way that they deviate from the field stars of Figure 1,
underscoring the comparability of the two samples.

To emphasize that the traditional distance to the Pleiades does not depend on
assumptions of age, in Figure 3
we show a CMD for nearby stars and the Pleiades, for $(m-M) = 5.6$.
The color used in Figure 3 is $(V-I)$ in the Cousins system, in order
to have an index that is less sensitive to metallicity than is \bv, and field
stars of all ages are represented.  The nearby
star parallaxes and colors are from the Hipparcos catalog, and we
used only stars with measured $(V-I)$, excluding those where $(V-I)$
had been estimated from \bv\ or other colors.
The Pleiades data are from Stauffer (1997, private
communication), who transformed his observations of Pleiads in the Kron
$(V-I)$ color (\cite{St84} 1984) to Cousins $(V-I)$ using the relation of
\cite{BW87} (1987),
correcting for reddening in the process.  The Pleiades $V$ magnitudes have
been shifted by 5.6 for distance and 0.12 to correct for extinction.
Both main sequences overlap for $(V-I) \la 1.7$.  The Pleiads redder than
this depart from the field star sequence simply because they are so
young that they lie above the main sequence.  There are essentially no nearby
stars below the ZAMS defined by the Pleiades.

\section{Sub-Luminous Stars}

We have just shown that nearby young stars lie on or above the usually
accepted ZAMS and that none lie below.  We now show that those stars that
do lie below the ZAMS are old stars of low metallicity, not young stars
analogous to Pleiads.

We began by extracting from the Hipparcos catalog all stars within 60 pc.
We kept only those stars with $\sigma_\pi/\pi < 0.050$ and $\sigma(B-V)
< 0.025$.  That portion of those stars that lie below the ZAMS is shown in
Figure 4.  We observed six of these stars, which are in squares in Figure 4.
We used the Hamilton spectrograph on the Lick 3 m Shane reflector, reducing the
data in the usual way within IRAF (see SSHJ for details).  The stars and the
spectroscopic results are listed in Table 2.  [Fe/H] was determined from the
strength of the \ion{Fe}{1} 6750 \AA\ line in comparison to a solar spectrum of
similar high resolution.  This is a small number of stars due to poor observing
conditions, but they were chosen randomly from the stars that lie about 0.3
magnitude below the ZAMS.

As we anticipated, most of these stars have unresolved rotation, and have
metallicities that are sub-solar, which accounts for their locations in the
CMD.  (\cite{Car94} 1994 show [Fe/H] = $-0.61$ for HIP 23431, in accord with
our value.)  There is one star, HIP 25127, that has obvious filling-in of the
H$\alpha$ line (Fig. 5).  This star also has relatively strong Li, the indicated
equivalent width implying $\log N{\rm (Li)}\approx 2.35$.  Also, we estimate
\vsini\ for HIP 25127 to be approximately 7 \kms, based on a comparison of
line breadths in this star to others in the sample.  All these factors suggest
youth, but this star's position in the CMD is due to its low metallicity of
$-0.3$, and so HIP 25127 validates models of ZAMS stars by confirming that
low-metals stars appear to lie well below the solar-metallicity ZAMS, even
if they may be young.

The symbols in Figure 4 indicate the transverse velocities of the
subluminous stars, calculated from the Hipparcos proper motions and
parallaxes.  Small filled circles indicate $v_{\rm trans} < 30$ \kms
(the median velocity for all stars within 50 pc).  Small circles indicate
$30 \leq v_{\rm trans} < 100$ \kms\ (the 95th percentile), while the large
circles have transverse velocities that exceed 100 \kms.  The scarcity of
low-velocity stars and the higher velocities of the more subluminous stars
strongly suggest that the objects in Figure 4 represent an old population,
lying below the ZAMS because of low metallicity.  The lack of subluminous
stars with \bv $\la 0.5$ is also indicative of an old population.

A more detailed examination of the kinematics of these stars requires
radial velocities to provide the third dimension, and a more strictly
limited sample to minimize the effects of observational errors.  For this
purpose, we extracted stars within 50 pc from the
Hipparcos catalog, accepting only those stars with $\sigma_\pi/\pi < 0.05$
and $\sigma(B-V) < 0.025$.   Binaries and stars with other astrometric
problems were rejected using flag H59 (ESA 1997, vol. 1, p. 126).
This left a clean sample with 3,345
stars.  Of these, we found radial velocities in the Hipparcos Input
Catalog for 1,799 of them, and these were used to calculate Galactic space
motions $U$, $V$, and $W$.  Correction to the Local Standard of Rest (LSR)
was done using the new solar motion $(U, V, W)_\odot^{\rm LSR} =
(+10, +5, +7)$ \kms\ from Hipparcos data (\cite{DB97} 1997).

Figure 6
shows the $(U, V)_{\rm LSR}$ and $(V, W)_{\rm LSR}$ diagrams for these 1,799
stars.  The sample has been divided into 1,598 stars lying on or above the
ZAMS (left panels) and 201 stars falling 0.1 or more magnitudes below the
ZAMS (right panels).  Table 3
summarizes the kinematic properties of these stars.  The net range of
velocities is roughly the same for both samples, but the ZAMS-and-above sample
is highly concentrated near the LSR, and its core shows vertex deviation and
clumpiness, which are characteristics of a young, low-velocity population.  By
contrast, these characteristics are completely absent in the diffuse velocity
distribution of the subluminous stars.  The conclusions to be drawn from
Figure 6 and Table 3 are clear: The 201 stars that appear to lie below the
ZAMS are dispersed in velocity space and chiefly represent the Galaxy's thick
disk population, with a small admixture of halo stars.  These stars are
subluminous simply because this old population has metallicities substantially
below solar.

It is still possible, of course, that some small fraction of these
subluminous stars are, in fact, young.  Such stars should have low space
motions; to attempt to identify any young, subluminous stars we selected 30
stars with \bv $<1.2$ that lie within 1 $\sigma$ of the LSR in all three
coordinates.  Ten of these stars, listed in Table 4,
fall 0.2 magnitude or more below the ZAMS.  We used SIMBAD to search for
additional information which might indicate the ages of these 10 stars or
reveal the reasons for their apparent subluminosity.  As indicated in
Table 4, most of these stars have low [Fe/H] or some other spectroscopic
indication of old age (such as weak H and K emission or a low Li
abundance).  Two stars appear to have significant errors in their colors,
including the only star of the 10 with any evidence of youth.

\section{Conclusions}

We have been unable to find any nearby stars with solar metallicity that are
very young and which are below the traditional ZAMS, despite the Hipparcos
results which suggest that Pleiades stars are 0.3 magnitude fainter than that
ZAMS.  We have also shown that those nearby stars that do lie below the ZAMS
show evidence of old age, as expected.

This leaves two possibilities.  The first is that the Hipparcos parallaxes for
the Pleiades and other clusters are correct but that Pleiades-like stars are
rare in the immediate solar neighborhood or we have just been unlucky in
finding them.  Surveys of activity in nearby stars have been comprehensive
enough to not have missed any significant number of genuinely young stars, and
we cannot accept the {\it ad hoc} explanation that the Pleiades is simply
bizarre.  We should note here that \cite{Gat95} (1995) measured a parallax
for the Coma cluster and found those stars to be subluminous to an extent
similar to what is found for the Pleiades.  However, Gatewood's result
depends on only three stars in Coma, one of which had especially large
uncertainty.  Moreover, the proper motions of those two remaining stars
differ significantly.  Thus Gatewood's measurements are intriguing but are
not sufficient to substantiate Hipparcos.  (Also, Gatewood's Coma results
differ from the traditional measures in a sense opposite to that seen by
Hipparcos, meaning that they conflict with each other.)

If stars in the Pleiades are indeed 0.3 magnitude fainter than we have
thought up to now, then there are significant aspects
of stellar physics that have so far gone unappreciated.  The companion
paper by \cite{Pins98} (1998) shows how difficult this notion is
to accept, but if this is true, then we surely cannot trust our inferences
of distances to the globular clusters if we cannot reproduce the behavior
of stars that are nearly identical to the Sun.

The second possibility is that the Hipparcos parallaxes have small
systematic errors.  The correction needed to bring the Hipparcos
Pleiades distance into agreement with the traditional value is almost
exactly one milliarcsec.  As shown in the companion paper by \cite{Pins98}
(1998), the Hipparcos parallaxes of the brightest Pleiads
in the core of the cluster are the most discrepant and weigh most heavily
in the net cluster parallax because of their low formal errors.  For this
reason we suspect that the Hipparcos net parallax for the Pleiades
is wrong.

The detection and measurement of visual binary orbits for Pleiades stars could
provide an independent estimate of the cluster's distance.  Such binaries
would be difficult to observe, but are within the capabilities of the Fine
Guidance Sensors on the {\it Hubble Space Telescope}, for example.

\acknowledgments
This work was supported, in part, by NASA grant NAGW-4837 to DS\@.  RBH and
BFJ acknowledge partial support from NASA Grant NAG5-4830 and NSF Grant
AST 9530632.  This research made use of the SIMBAD database, operated
by the CDS, Strasbourg, France.  We thank the anonymous referee for his or
her remarks.

\newpage

\figcaption{Absolute $V$ magnitude versus \bv\ color for nearby young
solar-type stars.  Parallaxes are from the Hipparcos output catalog,
while colors and magnitudes are from \cite{MM94} (1994).
The large solid circles represent
stars with high levels of chromospheric activity, taken to represent
the ZAMS.  The smaller solid circles are also active stars, but less so.
The open circles are active stars from the HK survey of \cite{Hen96} (1996).
The solid line is a theoretical
ZAMS from VandenBerg, calibrated as described in the text.  The dashed
line is the same ZAMS but uses the color-temperature relation of
\cite{Be79} (1979).  The diamond shows the position of the Sun.
For reference, the long-dashed line shows the ZAMS used in the companion
paper by Pinsonneault et al.}

\figcaption{The CMD for Pleiades solar-type stars.  The derivation of the
photometry is described in \cite{SSHJ} (1993) and is already
corrected for reddening of 0.04 magnitude in \bv\ and 0.12 magnitude
in $V$.  The lines are the same as in Figure 1, but displaced by a
distance modulus of 5.6 magnitudes.  The diamonds represent Pleiads that
are Ultra-Fast Rotators, meaning stars with \vsini $\ge30$ \kms.}

\figcaption{CMD for the Pleiades and for nearby stars, using Cousins
$(V-I)$ colors.  The colors and parallaxes for the nearby stars (small solid
circles) are from the Hipparcos catalog, and we have used only those stars for
which actual measured $(V-I)$ colors existed.  The Pleiades photometry (open
circles) is from Stauffer, and his Kron $(V-I)$ colors transformed to Cousins
colors using the \cite{BW87} (1987) relation.  The Pleiades colors have been
corrected by 0.06 magnitude for reddening.  The Pleiades magnitudes assume a
distance modulus of 5.6 and $A_V = 0.12$ magnitude.}

\figcaption{CMD for solar-type stars within 60 pc, with parallaxes good
to 10\% or better, and with $\sigma(B-V)<0.05$ magnitude.  Only those stars
lying below the ZAMS are shown.  The symbols denote different ranges of the
transverse velocity ($v_{\rm trans}$).  Dots have $v_{\rm trans}<30$ \kms,
small circles have $v_{\rm trans}$ from 30 to 100 \kms, and large circles
denote $v_{\rm trans}>100$ \kms.  The stars in squares were observed at high
resolution and are listed in Table 2.  The lines are the same as in Figure 2.}

\figcaption{The H$\alpha$ profile of HIP 25217 compared to
HIP 2128.  HIP 2128 shows the usual deep H$\alpha$ absorption profile of an
old, inactive star, while HIP 25217 exhibits significant filling-in of
H$\alpha$ due to chromospheric activity.  However, HIP 25217 has [Fe/H]
$\approx-0.3$.}

\figcaption{Galactic space motions for nearby stars ($d<50$ pc).  The starting
sample was taken from the Hipparcos catalog and includes 3,345 stars
with $\sigma_\pi/\pi < 0.05$ and $\sigma(B-V) < 0.025$, with binaries and
other problematic objects rejected.  Radial velocities were available for
1,799 stars from the Hipparcos Input Catalog, and those were used
with the parallax and proper motions to calculate $U$, $V$, and $W$ in a
right-handed coordinate system.  The motions have been corrected for the
solar motion relative to the Local Standard of Rest.  The left-hand
panels shows the 1,598 stars that lie on or above the ZAMS, while the
right-hand panels shows the 201 stars that lie $>0.1$ magnitude
below the ZAMS.}

\newpage

\begin{center}
\begin{deluxetable}{cccccccl}
\tablewidth{0pc}
\tablecaption{Absolute Magnitudes and Colors of Active Solar-Type Stars}
\tablehead{
\colhead{HIP}  & \colhead{HD} & \colhead{$(B-V)$}
 & \colhead{$V$} & \colhead{$\pi$ (mas)} & \colhead{$M_V$}
 & \colhead{$\log R^\prime_{\rm HK}$}
}
\startdata
\cutinhead{a) Most Active Stars}
   544 &    166 & 0.75  & 6.10 &  $72.98\pm0.75$ & 5.39 & $-4.33$  \nl
  8486 &  11131 & 0.61  & 6.75 &  $43.47\pm4.48$ & 4.91 & $-4.44$  \nl
 13402 &  17925 & 0.87  & 6.04 &  $96.33\pm0.77$ & 5.97 & $-4.30$  \nl
 25278 &  35296 & 0.53  & 4.99 &  $68.19\pm0.94$ & 4.17 & $-4.38$  \nl
 27913 &  39587 & 0.59  & 4.40 & $115.43\pm1.08$ & 4.70 & $-4.44$  \nl
 28954 &  41593 & 0.81  & 6.76 &  $64.71\pm0.91$ & 5.81 & $-4.42$  \nl
 42438 &  72905 & 0.62  & 5.63 &  $70.07\pm0.71$ & 4.86 & $-4.33$  \nl
 46843 &  82443 & 0.77  & 7.00 &  $56.35\pm0.89$ & 5.80 & $-4.20$  \nl
 54745 &  97334 & 0.60  & 6.40 &  $46.04\pm0.90$ & 4.73 & $-4.40$  \nl
 62512 & 111456 & 0.46  & 5.85 &  $41.39\pm3.20$ & 3.91 & $-4.43$  \nl
 64532 & 115043 & 0.60  & 6.83 &  $38.92\pm0.67$ & 4.77 & $-4.43$  \nl
 64792 & 115383 & 0.58  & 5.21 &  $55.71\pm0.85$ & 3.92 & $-4.33$  \nl
 71631 & 129333 & 0.61  & 7.54 &  $29.46\pm0.61$ & 4.95 & $-4.23$  \nl
 73869 & 134319 & 0.68  & 8.42 &  $22.59\pm0.68$ & 5.17 & $-4.33$  \nl
 88694 & 165185 & 0.62  & 5.94 &  $57.58\pm0.77$ & 4.74 & $-4.39$  \nl
 92919 & 175742 & 0.91  & 8.08 &  $46.64\pm1.03$ & 6.50 & $-4.21$  \nl
107350 & 206860 & 0.59  & 5.94 &  $54.37\pm0.85$ & 4.64 & $-4.42$  \nl
\cutinhead{b) Other Active Stars}
  1803 &   1835 & 0.66  & 6.39 &  $49.05\pm0.91$ & 4.84 & $-4.42$  \nl
 26779 &  37394 & 0.84  & 6.22 &  $81.69\pm0.83$ & 5.77 & $-4.43$  \nl
 29525 &  42807 & 0.66  & 6.44 &  $55.20\pm0.96$ & 5.14 & $-4.44$  \nl
 46580 &  82106 & 1.01  & 7.20 &  $78.87\pm1.02$ & 6.68 & $-4.43$  \nl
 66704 & 119124 & 0.53  & 6.33 &  $39.64\pm0.71$ & 4.30 & $-4.42$  \nl
 80337 & 147513 & 0.63  & 5.39 &  $77.69\pm0.86$ & 4.82 & $-4.43$  \nl
103859 & 200560 & 0.97  & 7.69 &  $51.65\pm0.72$ & 6.26 & $-4.43$  \nl
115331 & 220182 & 0.80  & 7.36 &  $45.63\pm0.83$ & 5.66 & $-4.41$  \nl
\cutinhead{c) Southern Stars from Henry et al.\ (1996)}
   490 &    105 & 0.595 & 7.51 &  $24.85\pm0.92$ & 4.49 & $-4.36$ \nl
 14007 &  18809 & 0.677 & 8.47 &  $21.21\pm0.88$ & 5.11 & $-4.33$ \nl
 26990 &  38397 & 0.586 & 8.18 &  $19.17\pm0.73$ & 4.55 & $-4.26$ \nl
 28764 &  41700 & 0.517 & 6.35 &  $37.46\pm0.50$ & 4.22 & $-4.35$ \nl
 30001 &  44135 & 0.632 & 8.14 &  $16.29\pm0.88$ & 4.20 & $-4.33$ \nl
 36948 &  61005 & 0.734 & 8.20 &  $28.95\pm0.92$ & 5.54 & $-4.26$ \nl
 36832 &  61033 & 0.724 & 7.59 &  $35.27\pm0.65$ & 5.33 & $-4.34$ \nl
 37563 &  62850 & 0.637 & 7.20 &  $30.07\pm0.56$ & 4.56 & $-4.30$ \nl
 42808 &  74576 & 0.917 & 6.56 &  $89.78\pm0.56$ & 6.35 & $-4.31$ \nl
 43290 &  75519 & 0.651 & 7.98 &  $27.71\pm0.70$ & 5.04 & $-4.37$ \nl
 59315 & 105690 & 0.707 & 8.18 &  $26.43\pm1.03$ & 5.27 & $-4.27$ \nl
 63862 & 113553 & 0.678 & 7.90 &  $22.11\pm1.15$ & 5.05 & $-4.30$ \nl
 66765 & 118972 & 0.855 & 6.93 &  $64.08\pm0.81$ & 5.95 & $-4.39$ \nl
 69781 & 124784 & 0.650 & 8.77 &  $15.36\pm1.12$ & 4.60 & $-4.23$ \nl
 78505 & 142033 & 0.657 & 8.00 &  $17.22\pm0.79$ & 4.20 & $-4.35$ \nl
 82431 & 151598 & 0.673 & 8.23 &  $13.42\pm1.12$ & 3.87 & $-4.40$ \nl
 95149 & 181321 & 0.628 & 6.48 &  $47.95\pm1.28$ & 4.88 & $-4.31$ \nl
 96334 & 183414 & 0.648 & 7.92 &  $28.22\pm0.98$ & 5.14 & $-4.23$ \nl
 98704 & 188480 & 0.535 & 8.22 &  $12.05\pm0.98$ & 3.64 & $-4.35$ \nl
 98839 & 190102 & 0.626 & 8.18 &  $21.56\pm1.08$ & 4.83 & $-4.39$ \nl
 99137 & 190422 & 0.530 & 6.25 &  $43.08\pm0.79$ & 4.43 & $-4.38$ \nl
101432 & 195521 & 0.666 & 6.80 &  $25.48\pm0.89$ & 3.83 & $-4.31$ \nl
105612 & 202732 & 0.687 & 7.88 &  $29.21\pm0.75$ & 5.21 & $-4.40$ \nl
105384 & 203019 & 0.687 & 7.84 &  $27.49\pm1.18$ & 5.04 & $-4.36$ \nl
105712 & 203244 & 0.723 & 6.97 & $140.88\pm3.09$ & 5.43 & $-4.39$ \nl
113579 & 217343 & 0.655 & 7.48 &  $31.22\pm0.93$ & 4.94 & $-4.27$ \nl
117596 & 223537 & 0.666 & 8.03 &  $18.64\pm0.64$ & 4.38 & $-4.36$ \nl
\enddata
\end{deluxetable}
\end{center}

\begin{center}
\begin{deluxetable}{rrccccccc}
\tablewidth{0pc}
\tablecaption{Spectroscopic Observations of Six Subluminous Stars}
\tablehead{
\colhead{HIP}  & \colhead{HD} & \colhead{$(B-V)$} & \colhead{$V$}
 & \colhead{$\pi$ (mas)} & \colhead{$M_V$}
 & \colhead{$W_\lambda {\rm (Li)}$} & \colhead{[Fe/H]} & \colhead{$v_{\rm trans}$}
}
\startdata
     7 &    --- & $0.740\pm0.020$  & 9.64 &  $17.74\pm1.30$ & 5.88 & $\le3$ & $-0.31$ &  77 \nl
  2128 &    --- & $0.703\pm0.041$  & 9.66 &  $17.16\pm1.47$ & 5.83 & $\le3$ & $-0.68$ & 103 \nl
  5697 &   7320 & $0.685\pm0.003$  & 8.95 &  $21.71\pm1.25$ & 5.63 &     7  & $-0.50$ &  47 \nl
  7687 &  10166 & $0.794\pm0.001$  & 9.37 &  $25.82\pm1.56$ & 6.43 & $\le3$ & $-0.43$ &  49 \nl
 23431 &  32237 & $0.720\pm0.015$  & 8.19 &  $34.88\pm1.46$ & 5.90 & $\le3$ & $-0.64$ &  56 \nl
 25127 & 243086 & $0.706\pm0.029$  & 9.15 &  $20.54\pm1.22$ & 5.71 &    94  & $-0.27$ &  11 \nl
\enddata
\end{deluxetable}
\end{center}

\begin{center}
\begin{deluxetable}{lrrrrrrr}
\tablewidth{0pc}
\tablecaption{Kinematics of Nearby Stars}
\tablehead{
\colhead{Sample}  & \colhead{$N_{\rm stars}$} & 
 \colhead{$\langle U_{\rm LSR} \rangle$} &
 \colhead{$\langle V_{\rm LSR} \rangle$} &
 \colhead{$\langle W_{\rm LSR} \rangle$} &
 \colhead{$\sigma(U)$} &
 \colhead{$\sigma(V)$} & \colhead{$\sigma(W)$}
}
\startdata
All stars             & 1799 &  $-2.3$ & $-18.4$ & $-1.1$ &  40 &  31 & 19 \nl
$>0.1$ mag below ZAMS & 201  &   $0.0$ & $-26.4$ & $-1.1$ &  63 &  55 & 27 \nl
0.1 to 0.2 below ZAMS &  63  &  $-8.2$ & $-18.0$ & $+4.1$ &  43 &  33 & 20 \nl
0.2 to 0.5 below ZAMS & 108  &  $-2.8$ & $-18.4$ & $-2.4$ &  56 &  37 & 27 \nl
$>0.5$ below ZAMS     &  30  & $+27.4$ & $-72.5$ & $-7.6$ & 104 & 104 & 38 \nl
\enddata
\end{deluxetable}
\end{center}

\begin{center}
\begin{deluxetable}{rrcccc}
\tablewidth{0pc}
\tablecaption{Ten Low-Velocity Subluminous Stars}
\tablehead{
\colhead{HIP}  & \colhead{HD} & \colhead{$(B-V)$} &
 \colhead{$M_V$} & \colhead{Sp.} & \colhead{Note}
}
\startdata
  8275 &  10853   & 1.044  & $7.10\pm0.07$ & K5   & 1 \nl
 12184 &  16232   & 0.51   & $4.12\pm0.09$ & F4V  & 2 \nl
 44984 &  78661   & 0.355  & $3.60\pm0.07$ & F2p  & 3 \nl
 56998 & 101581   & 1.064  & $7.28\pm0.02$ & K5V  & 4 \nl
 58863 & 104828   & 1.072  & $7.35\pm0.11$ & K0   & 5 \nl
 68796 & 123710   & 0.590  & $5.00\pm0.06$ & G5   & 6 \nl
 80366 & 147776   & 0.950  & $6.73\pm0.06$ & K2V  & 7 \nl
 81520 & 149612   & 0.616  & $5.33\pm0.04$ & G3V  & 8 \nl
106231 & +22 4409 & 1.050  & $7.24\pm0.06$ & K8   & 9 \nl
114790 & 219249   & 0.695  & $5.54\pm0.07$ & G6V  & 8 \nl
\enddata
\tablenotetext{1}{Gliese 74, dK8 (\cite{GJ91} 1991).  Marginally subluminous.}
\tablenotetext{2}{Hipparcos \bv\ error; SIMBAD lists \bv=0.51, from \cite{Tol64} (1964).}
\tablenotetext{3}{[Fe/H]$=-0.3$ (\cite{CC71} 1971).}
\tablenotetext{4}{Gliese 435; \cite{Fav97} (1997) notes low Li.}
\tablenotetext{5}{\cite{Egg85} (1985) lists this star as a member of his
Hyades Supercluster, with $(m-M) = 2.82$ implying $M_V = 7.03$, which is
$3\sigma$ brighter than the Hipparcos $M_V$.}
\tablenotetext{6}{[Fe/H]$=-0.58$ (\cite{Cay97} 1997).}
\tablenotetext{7}{Gliese 621.  \cite{Egg86} (1986) lists this star as a member
of the 4 Gyr-old Wolf 630 Group.}
\tablenotetext{8}{Inactive star (\cite{Hen96} 1996).}
\tablenotetext{9}{LO Peg; $P=0.4$ day, $A=0.09$ mag (\cite{ESA97} 1997);
Hipparcos color is discrepant with spectral type of K8 (\bv $\sim1.2$).
\cite{Ster97} (1997) note high x-ray flux and strong Li feature, suggesting
youth.}
\end{deluxetable}
\end{center}

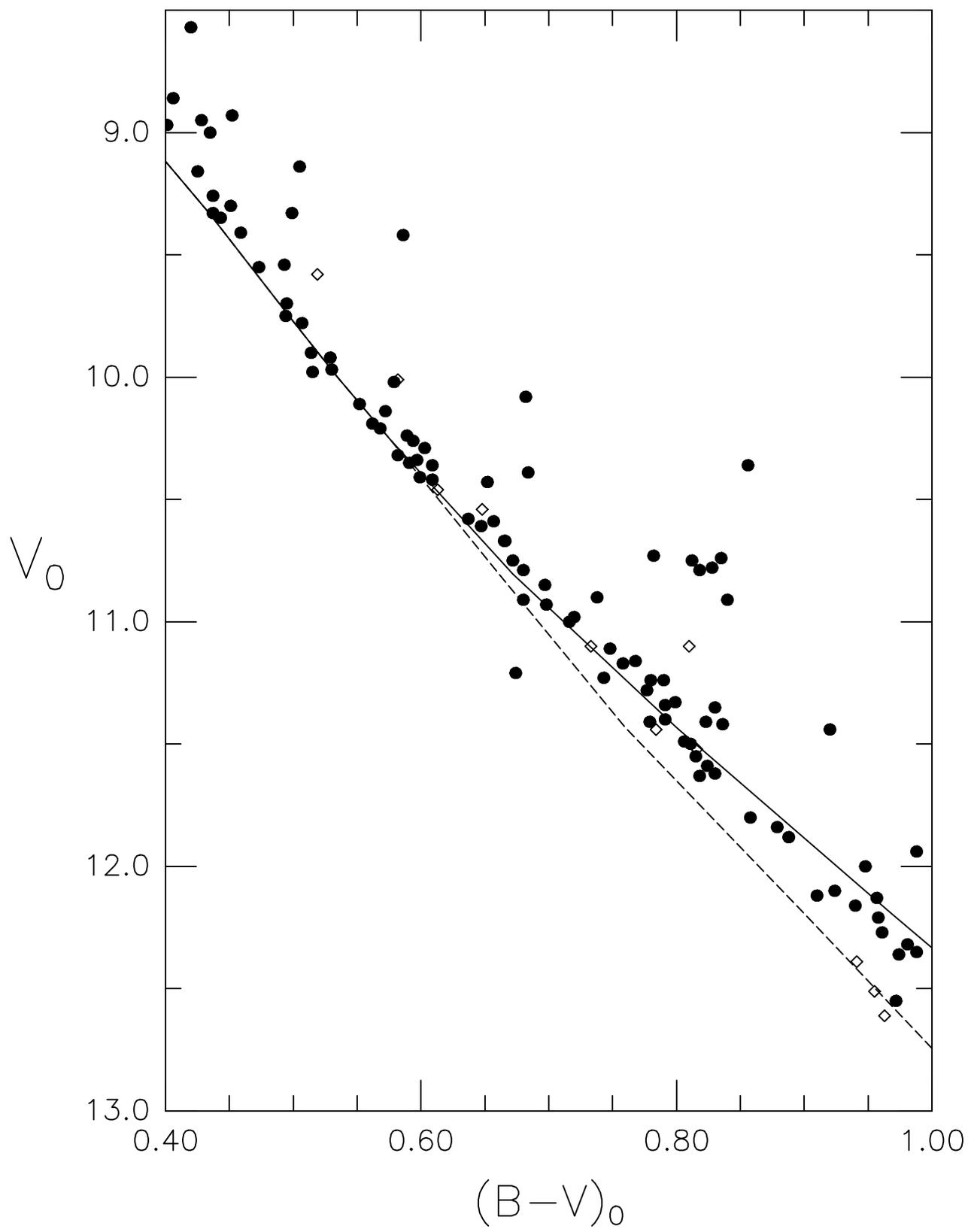
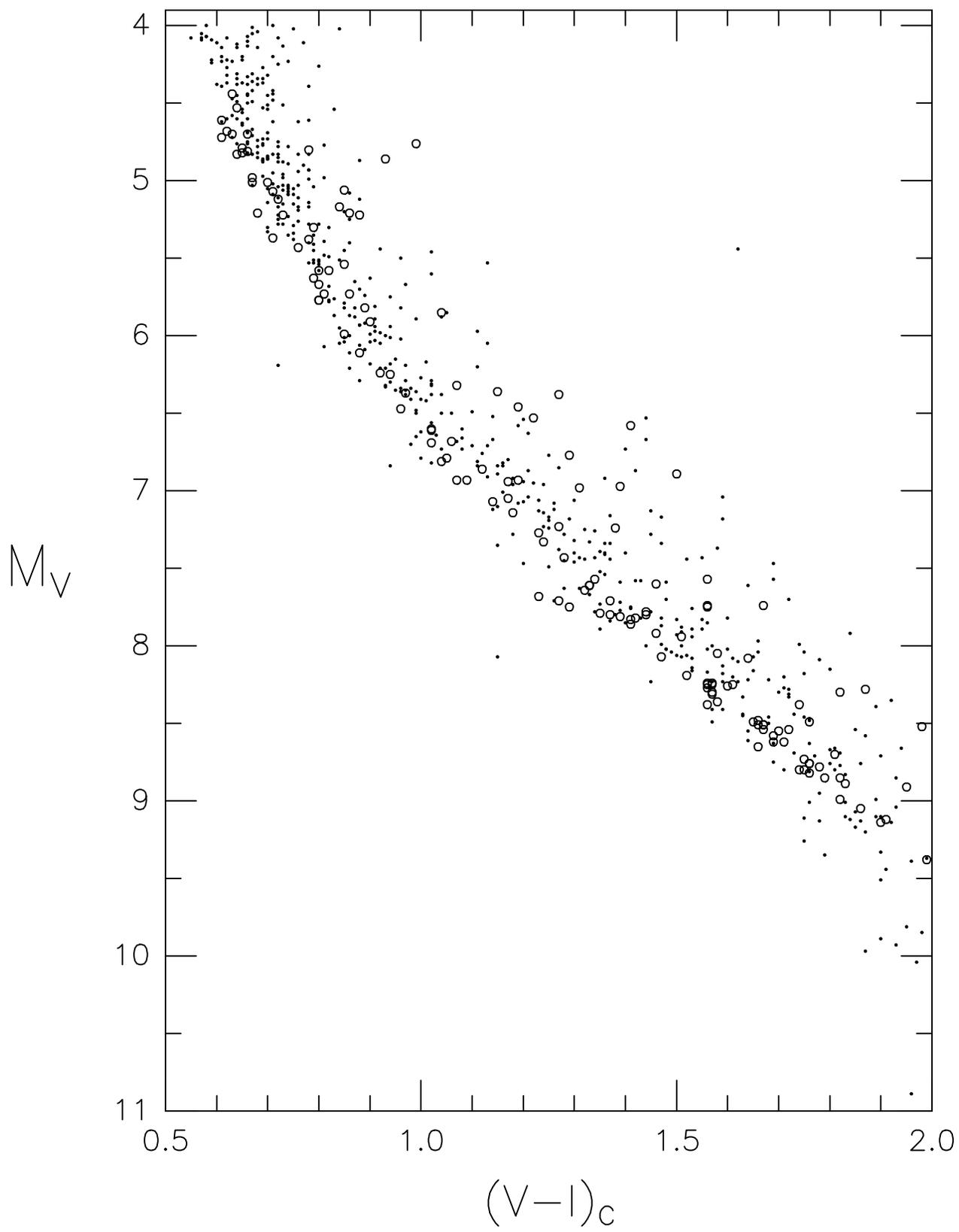
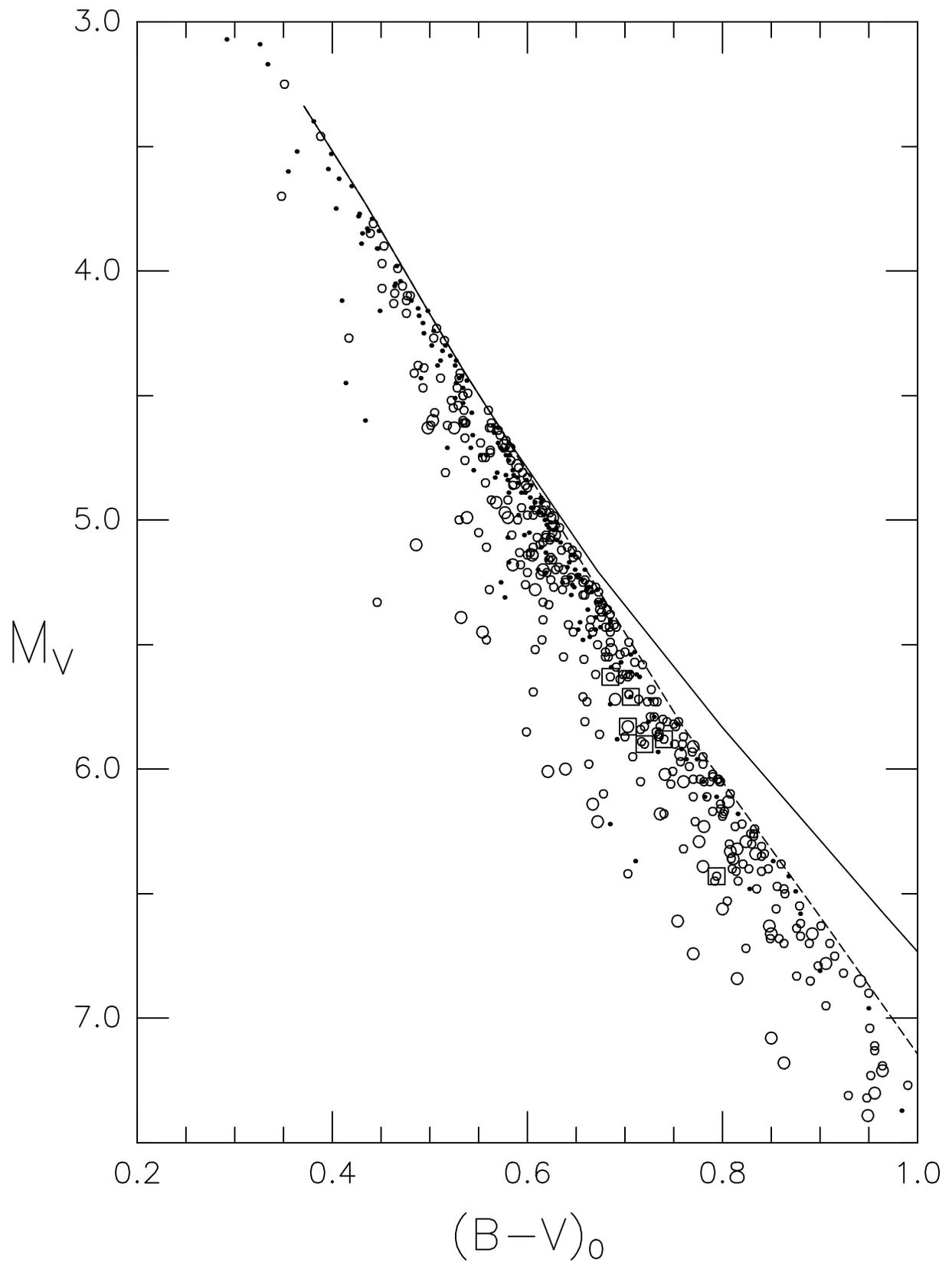
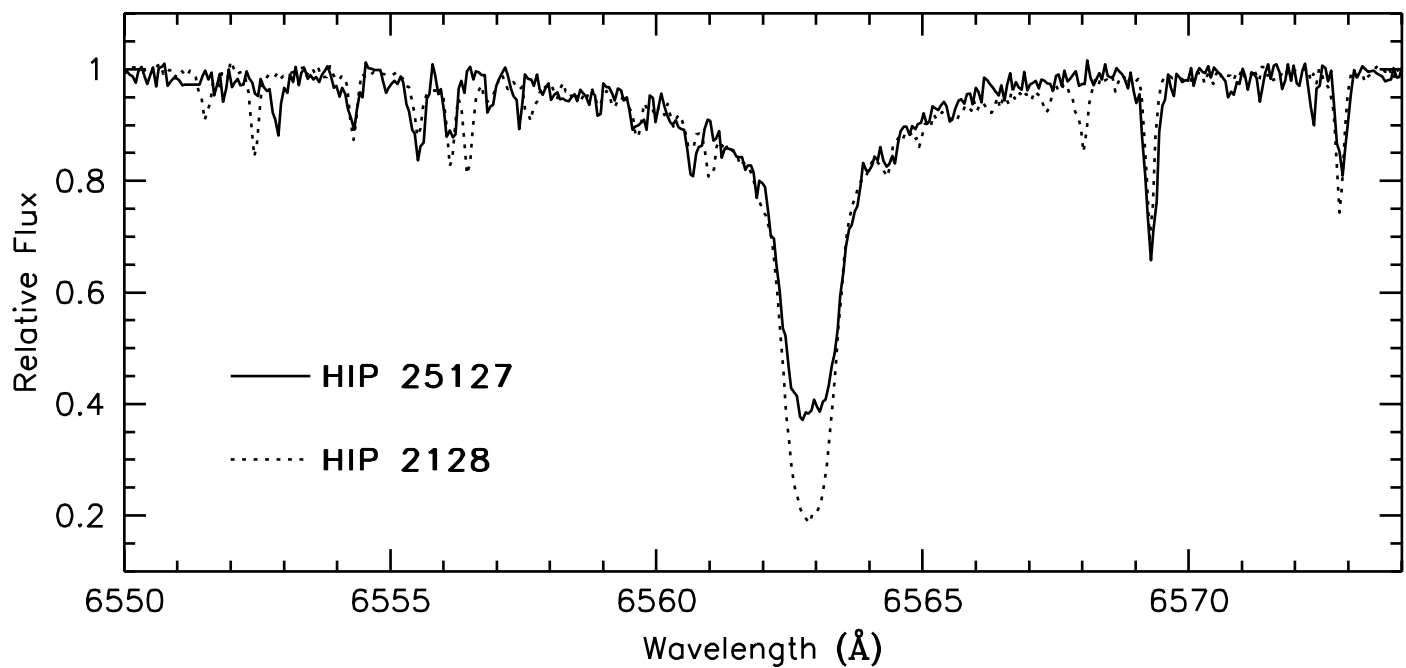
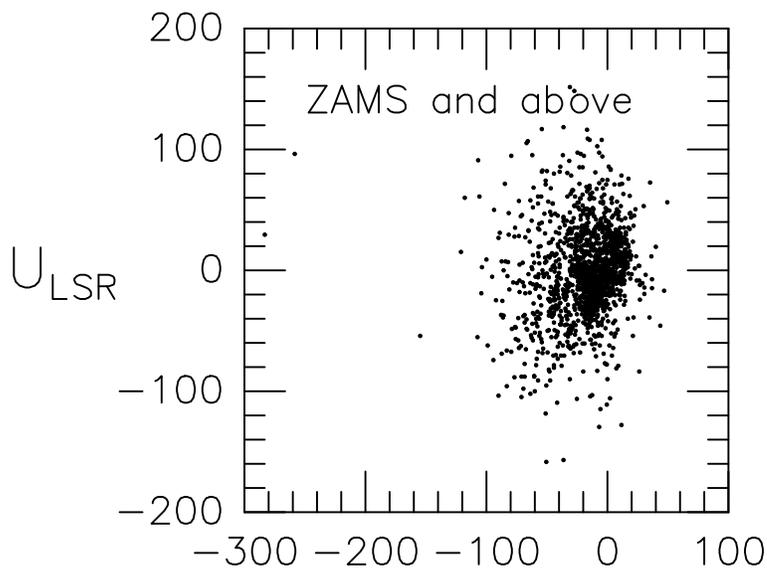
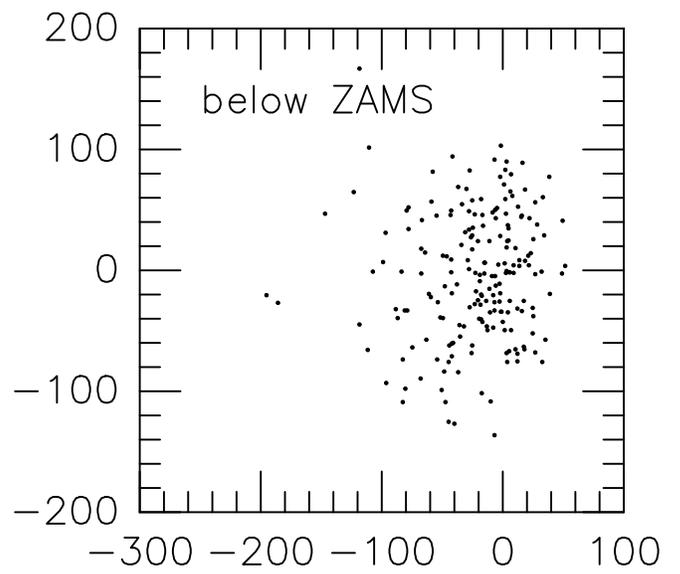
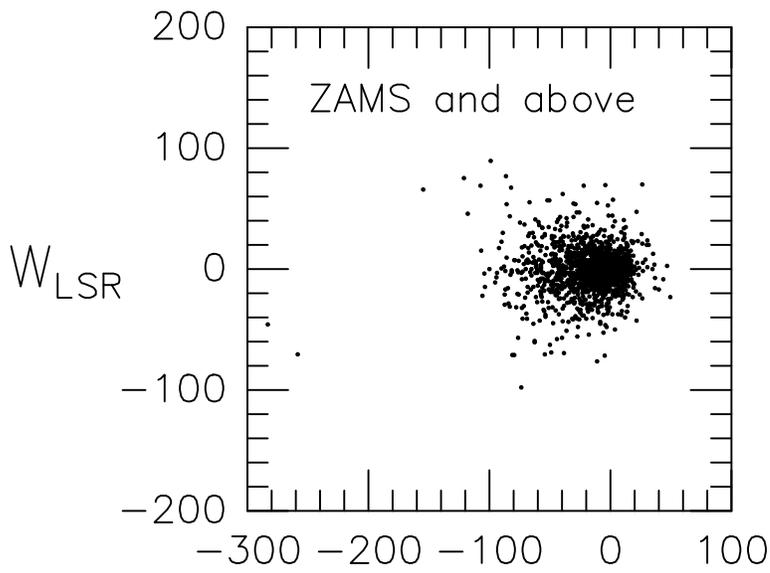
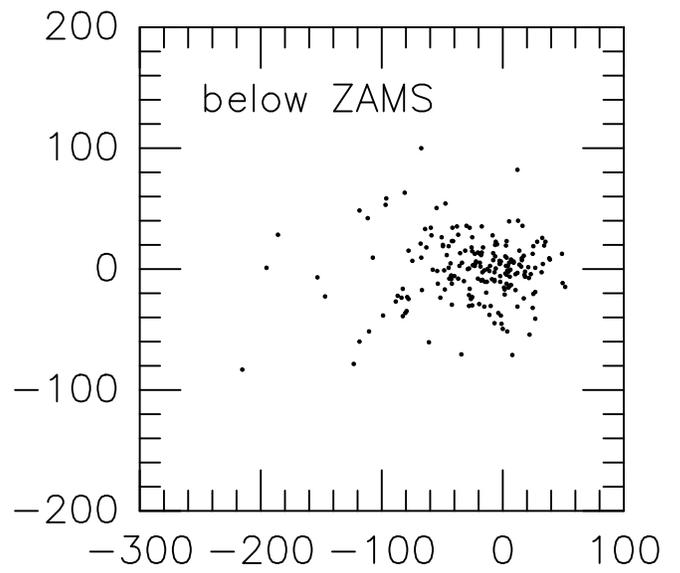
\end{document}